\documentclass[a4paper,12pt]{article}

\usepackage{latexsym,amsmath,amsfonts,amssymb}
\usepackage{multirow}
\usepackage[dvips]{graphicx}
\usepackage{mathrsfs,mathtools}
\usepackage{pstricks,pst-node,pst-text,pst-3d}
\usepackage{slashed}
\usepackage{verbatim}

\newcommand{\tr}[1]{\text{Tr}\big[#1\big]}

\newcommand{\sect}[1]{\setcounter{equation}{0}\section{#1}}

\begin{document}

\begin{titlepage}

\setcounter{page}{0}

\begin{flushright}
{ }
\end{flushright}

\vspace{0.6cm}

\begin{center}
  {\Large \bf Two and three-point correlators of operators dual to folded string solutions at strong coupling}

\vskip 0.8cm

{\bf George Georgiou}\footnote{georgiou@inp.demokritos.gr}
\\
{\sl
Demokritos National Research Center \\ Institute of Nuclear Physics\\
Ag. Paraskevi, GR-15310 Athens, Greece}\\

\vskip 1.2cm

\end{center}

\begin{abstract}
A particular analytic continuation of classical string solutions
having a single $AdS_5$ spin is considered.
These solutions describe strings tunnelling from the boundary to the boundary of $AdS_5$.
We use the Legendre transform of the dimensionally regularised action of these solutions to evaluate the 2-point functions of the dual operators,
holographically.
Subsequently, we evaluate
the structure coefficient of correlators involving
two operators with spin $S$ and a BPS state, at strong coupling.
Our expressions are valid for any value of the
AdS spin $S/\sqrt{\lambda}$ and can be applied both at the  case of long and short strings.
For long strings and at leading order, the structure coefficient is independent of the spin $S$ for twist J operators,
while it scales as $1/\log^l{\frac{S}{\sqrt{\lambda}}}$ for the case of operators with two equal
angular momenta in $S^5$.
For short strings, the structure coefficient is proportional to the energy $E$ of the string.
Finally, we comment on the possibility of relating
the strong coupling 3-point function coefficient of three large spin twist 2 operators
to the action of the 6-gluon scattering amplitude.
\end{abstract}

\vfill

\end{titlepage}

\sect{Introduction}\label{sec:intro}

Conformal Field Theories (CFTs) are partially characterised
by the identification of their primary
operators and the knowledge of the conformal dimensions of these operators.
The second
crucial characterisation of a CFT is given by the structure constants
which determine the Operator Product Expansion (OPE) between two
primary operators.  ${\cal N}=4$ Super Yang-Mills  (SYM) theory is an
important example of an interacting four dimensional CFT which has been
thoroughly studied because of the AdS/CFT duality with string
theory~\cite{Maldacena:1997re,Witten:1998qj}. In particular, in the recent years huge
progress has been made in the computation of the planar contribution
to the conformal dimensions of non-protected
operators for any value of the coupling constant,
using integrability. On the other hand, very little is known about the
structure constants.

Conformal symmetry determines the 3-point functions up to an overall
coupling-dependent constant $C_{123}$. In the simple case of scalar primary operators
their 3-point function takes the form
\begin{eqnarray}\label{3pointconformal}
<{\cal O}_1(x_1)\,{\cal O}_2(x_2)\,{\cal O}_3(x_3)>=\frac{C_{123}}{
|x_{12}|^{\Delta_1+\Delta_2-\Delta_3}\,|x_{23}|^{\Delta_2+\Delta_3-\Delta_1}\,
|x_{13}|^{\Delta_1+\Delta_3-\Delta_2}},
\end{eqnarray}
where $\Delta_i, \,i=1,2,3$ is the dimension of the operator ${\cal O}_i$.
In order to be able to define the structure constant $C_{123}$
unambiguously, it is important to normalised the 2-point functions of  ${\cal O}_i$
to one. Namely,
\begin{eqnarray}\label{2pointconformal}
<{\cal O}_i(x_1)\,{\cal O}_j(x_2)>=\frac{\delta_{ij}}{x_{12}^{2\Delta_i}}.
\end{eqnarray}

In principle it is possible to tackle the problem of finding $C_{123}$
both at  weak and at strong coupling. In the
first case the structure constants are extracted from the 3-point
correlators among gauge invariant operators, while in the second
case one needs to compute the partition function of IIB string
theory in AdS$_5 \times S^5$ with appropriate boundary
conditions~\cite{Gubser:1998bc,Witten:1998qj}. Unfortunately neither
of these approaches can be currently used to explicitly evaluate the
structure constants as an exact function of 't~Hooft coupling
($\lambda$) even in the planar limit.

Our current knowledge of the OPE coefficients is essentially based on
a perturbative expansion around $\lambda=0$, where standard Feynman
diagrams can be used to evaluate the relevant gauge theory
correlators, or around $\lambda=\infty$ where the IIB string theory is
well approximated by a simpler description.  By comparing the 3-point
correlators among half-BPS operators in these two different limits,
the authors of~\cite{Lee:1998bx} conjectured that the corresponding
structure constants are non-renormalised (i.e. they have a trivial
dependence on the 't~Hooft coupling).  On the contrary the 3-point
correlators among non-protected operators receive quantum corrections,
as it is shown, for instance, by the correlator between three Konishi
operators~\cite{Bianchi:2001cm}. On the gauge theory side, the authors
of~\cite{Okuyama:2004bd,Roiban:2004va,Alday:2005nd,Alday:2005kq,Grossardt:2010xq}
studied systematically the structure constants for operators with only
bosonic fields and computed the corrections arising from the planar
1-loop Feynman diagrams. The importance of the operator mixing for the operators
participating in the correlators was stressed in \cite{Georgiou:2008vk,Georgiou:2009tp,us}.
 On the string theory side it is more
difficult to extract information about non-protected OPE coefficients,
since, in the supergravity limit, all non-protected operators acquire
large conformal dimension and decouple. The BMN
limit~\cite{Berenstein:2002jq} represents a different approximation,
where it is possible to extract useful information on non-BPS
structure constants.

Recently, another approach to the calculation of n-points correlators
involving non-BPS states was developed
\cite{Tsuji:2006zn,Janik:2010gc,Buchbinder:2010vw,Zarembo:2010rr,Costa:2010rz,us2}.
More precisely, the authors of \cite{Janik:2010gc} argued that it should be possible to
obtain the correlation functions of local operators corresponding to classical
spinning string states, at strong coupling, by evaluating the string action on
a classical solution with appropriate boundary conditions
after convoluting  with the relevant to the classical states wavefunctions.
In \cite{Buchbinder:2010vw}, the 2-point correlator of vertex operators representing
classical string states with large $AdS_5$ spin was computed and agreement with the 2-point
function of twist 2 operators, at strong coupling, was obtained.
Finally, in \cite{Zarembo:2010rr,Costa:2010rz,Roiban:2010fe,Hernandez:2010tg,Ryang:2010bn,Arnaudov:2010kk} the 3-point function coefficients
of a correlator involving a massive string state, its conjugate and a supergravity
state was computed. This was done by taking advantage of the known classical solutions
corresponding to the 2-point correlators of operators dual to massive string states.

The plan for the rest of this paper is as follows.
In Section 2.1, we consider a particular analytic continuation of
the Gubser-Klebanov-Polyakov (GKP) folded string \cite{Gubser:2002tv}
which is believed to be the gravity dual of the large spin twist 2
operators. The analytically continued solution describes the propagation of a string
extending along a light-like direction on the boundary of the $AdS_5$
into the bulk and back to the boundary. We use the action of this solution
to calculate the 2-point function of twist 2 operators at strong coupling, holographically.
To achieve this we have to perform a Legendre transform of the dimensionally regularised
action with respect to the angle conjugate to the spin of the string $S$.
In Section 2.2 we apply the aforementioned method to obtain the 2-point function of twist J operators,
while in Section 2.3 the 2-point function of operators dual to string states with one large spin
in $AdS_5$ and two large and equal spins in $S^5$ is calculated.
In all cases, both the spacetime structure and the scaling dimension of the operators is correctly recovered.
In Section 3, we use the formalism of \cite{Zarembo:2010rr} to evaluate
the 3-point function coefficient, at strong coupling, of a correlator involving
the two operators with spin $S$ considered in Sections 2.1, 2.2 and 2.3, and a BPS state.
Again the role of the solutions written in Section 2 is crucial.
The 3-point correlator is essentially obtained from a fattened Witten diagram
where the supergravity state propagates from a point on the boundary of $AdS$
to a point on the worldsheet of the classical solutions describing the tunneling
strings of Section 2 \cite{Zarembo:2010rr}.
Our expressions for the two and three point functions are valid for any value of the
AdS spin $S/\sqrt{\lambda}$ and can be applied both at the  case of long and short strings.
For long strings, $S/\sqrt{\lambda}>>1$, the structure coefficient appears to be proportional to
$\sqrt{\lambda}$ and independent of the spin $S$ at leading order for the twist J operators,
while it scales as $1/\log^l{\frac{S}{\sqrt{\lambda}}}$ for the case of the operators with two equal
angular momenta in $S^5$. Here $l$ is characterising the BPS state which transforms in the
$[0,l,0]$ representation of $SU(4)$.
For short strings, $S/\sqrt{\lambda}<<1$, the structure coefficient is proportional to
$\sqrt{2S \sqrt{\lambda}}=E$, which is the energy of the short string state.

We close Section 3 by commenting on the
possibility of obtaining the strong coupling 3-point function coefficient of three twist 2 operators
from the minimal surface of the six gluon scattering amplitude in $N=4$ SYM.

\sect{2-point functions of operators with a single spin in $AdS$}
\label{sec:2-pointgen}
In this Section, we consider a particular analytic continuation of classical string solutions
having a large $AdS_5$ spin. The continuation is done both at the worldsheet time, as well as,
at some of the target space coordinates in such a way that the $AdS$ coordinates of the solution remain real. Furthermore,
the signatute of the target space remains Minkowskian.
Thus one can have a nice spacetime interpretation of the solution. As we will see,
it appears as a string tunneling from the boundary to the boundary of the $AdS$ space. This is to be
contrasted with the solution we start from which lives entirely in the bulk of $AdS$ and never touches the boundary.
Then we use this solution to evaluate the 2-point functions of the dual operators holographically.
This is achieved by performing a  series of Legendre transforms of the corresponding
action with respect to the angles conjugate to the angular momenta of the string.

\subsection{2-point function of twist 2 operators}
\label{sec:2-point}

The anomalous dimension of twist 2 operators
has been studied extensively, both at weak coupling \cite{Callan,Gross,Axenides:2002zf,Kotikov:2004er}
for theories with different amounts of supersymmetry
and at
strong coupling \cite{Gubser:2002tv,Frolov:2002av,Kruczenski:2002fb,Basso:2007wd,Basso:2006nk,Kruczenski:2007cy},
for the maximally supersymmetric theory.
In this Section, we will calculate the 2-point function of twist 2
operators at strong coupling, by means of the AdS/CFT correspondence.
This will be achieved by writing down a solution of the Polyakov action that describes
a string which tunnels from the boundary to the boundary of the $AdS_5$ space.
Once we perform a Legendre transform of the classical action
of this solution, it is possible to get the dimension as well as the spacetime
dependence of the 2-point function, as this is dictated by conformal invariance.

The starting point is the Gubser-Klebanov-Polyakov (GKP) solution \cite{Gubser:2002tv}
which has been argued to be the strong coupling dual of twist 2 operators
with large spin $S$. We scematically write these operators as
${\cal O}_S=\tr {\Phi D^S_+ \Phi}+...$, where $\Phi$ is a scalar field,
e.g. in $N=4$ SYM it can be thought as one of the six scalars of the
$N=4$ supermultiplet and $D_+$ is a covariant derivative
along a light-cone direction. The dots stand for mixing terms having
fermions or gluons instead of scalars as well as for terms where the
derivative are distributed differently among the two scalars.

The GKP solution written in global coordinates reads:
\begin{equation}\label{GKP}
t=k \tau, \, \phi=w t, \, \rho=\rho(\sigma), \, \rho \in [0,\rho_0],\, \sigma \in [0,2 \pi]
\end{equation}
where $\phi$ is a maximal circle of the $S^3$ of $AdS_5$
, $\frac{w^2}{k^2}=\cosh^2{\rho_0}/\sinh^2{\rho_0}$ and $\rho(\sigma)$ is the solution of the
equation
\begin{equation}\label{rho'}
\rho'^2=k^2
\cosh^2{\rho}-w^2\sinh^2{\rho}.
\end{equation}
\eqref{GKP} describes a folded closed string which rotates at the centre
of $AdS_5$ with spin $S=2/\pi \, \sqrt{\lambda}\,w\int_{0}^{\pi/2}d\sigma \sinh^2{\rho} $ and energy
$E=2/\pi\, \sqrt{\lambda}\,k\int_{0}^{\pi/2}d\sigma \cosh^2{\rho} $.
Its tips,which are at $\rho_0$, approach the boundary of the $AdS$ space, which in
global coordinates is at $\rho=\infty$, as the spin scales
to infinity $S \rightarrow \infty$.
One can calculate the energy of this solution and verify that
for large enough spin it is given by the expression
\begin{equation}\label{EGKP}
E=S+\frac{\sqrt{\lambda}}{\pi} \log{\frac{S}{\sqrt{\lambda}}}.
\end{equation}
On general grounds, one expects to find an expression of the
form
\begin{equation}
E=S+f(\lambda)\log{\frac{S}{\sqrt{\lambda}}} +O(S^0),
\end{equation}
where $f$ is the so-called cusp anomalous dimension which
can be expanded either around weak or strong coupling.
On the gauge theory side, $E$ should be interpreted as the dimension $\Delta$
of a large spin twist 2 operator, according to AdS/CFT.
From \eqref{EGKP} one can immediately read the first term
in the strong coupling expansion,
namely $f(\lambda)=\frac{\sqrt{\lambda}}{\pi}+...$.
In general, it is believed that $f(\lambda)$
is a smooth interpolating function
connecting the two perturbative expansions.
Indeed, assuming that $N=4$ SYM is integrable, one can, in principle,
compute as many terms in the expansion around  both weak and strong coupling
as one wishes.

Although the solution \eqref{GKP} is perfectly fine for calculating
the cusp anomalous dimension, it is not appropriate for calculating
the 2-point function of twist 2 operators holographically.
This is because the string of \eqref{GKP} lives entirely in the
bulk of $AdS$ and it never touches its boundary.
What we need is a string solution that tunnels from the boundary
of $AdS_5$ to the boundary, in the spirit of \cite{Yoneya:2006td,Dobashi:2004nm}.

This solution can be constructed from \eqref{GKP} by rewriting
it in embedding coordinates:
\begin{eqnarray}\label{GKP2}
Y^{-1}=\cos{t}\cosh{\rho},  \qquad Y^0=\sin{t} \cosh{\rho},
\qquad Y^1=\cos{\phi}\sinh{\rho}, \nonumber \\
\qquad Y^4=\sin{\phi}\sinh{\rho}, \qquad Y^2=Y^3=0,
\qquad \rho=\rho(\sigma),\, \phi=w \tau, \,t= k \tau.
\end{eqnarray}
Obviously, \eqref{GKP2} can be embedded in a $AdS_3$ subspace of
the full $AdS_5$.
We can now analytically continue the world-sheet time
$\tau \rightarrow i \tau$ while at the same time we perform
an analytic continuation to the embedding coordinates of the
form \cite{Alday:2010ku}:
\begin{equation}\label{continuation}
Y^0\rightarrow i Y^4, \qquad Y^4\rightarrow i Y^0,\qquad
\tau \rightarrow i \tau,\qquad \phi\rightarrow i \phi.
\end{equation}
Taking into account \eqref{continuation}, \eqref{GKP2} yields:
\begin{eqnarray}\label{GKP3}
Y^{-1}=\cosh{t}\cosh{\rho},  \qquad Y^0=\sinh{\phi} \sinh{\rho},
\qquad Y^1=\cosh{\phi}\sinh{\rho}, \nonumber \\
\qquad Y^4=\sinh{t}\cosh{\rho}, \qquad Y^2=Y^3=0,
\qquad \rho=\rho(\sigma),\,t= k \tau ,\, \phi=w \tau.
\end{eqnarray}
Let us now comment on the solution \eqref{GKP3}.
Firstly, we would like to note that because of the double
Wick rotation of \eqref{continuation}, the target spacetime
is defined by $-(Y^{-1})^2-(Y^{0})^2+(Y^{1})^2+(Y^{2})^2+(Y^{3})^2+
(Y^{4})^2=-1$ and as a consequence it remains Minkowskian.
This is to be contrasted to the construction of \cite{Yoneya:2006td,Dobashi:2004nm,Tsuji:2006zn}
where it is essential that the target space is Euclidean $AdS_5$.\\
Secondly, it is easy to see that \eqref{GKP3}
describes a string which tunnels from the boundary to the
boundary of the $AdS_5$ space. Indeed, at $\tau=-\infty$
\eqref{GKP3} directly gives $Y^4=-Y^{-1}$ and $Y^0=-Y^{1}$ while
at $\tau=\infty$ it gives $Y^4=Y^{-1}$ and $Y^0=Y^{1}$.
Taking into account that the boundary of $AdS_5$ is defined by
the relations $Y^2=0, Y^{\mu}\sim c\, Y^{\mu}$ we conclude that
we have a string extending along a light-like direction on the boundary at $\tau=-\infty$ that
propagates in the bulk to end on the boundary again at $\tau=\infty$.
Furthermore, in the limit where the spin $S$, and as
a consequence $\rho_0$ tends to infinity, the tips of the
string touch the boundary writing light-like
segments \cite {Buchbinder:2010vw,Alday:2010ku}.
\begin{figure}[!t]
  \centering
  \includegraphics[width=.30\textwidth]{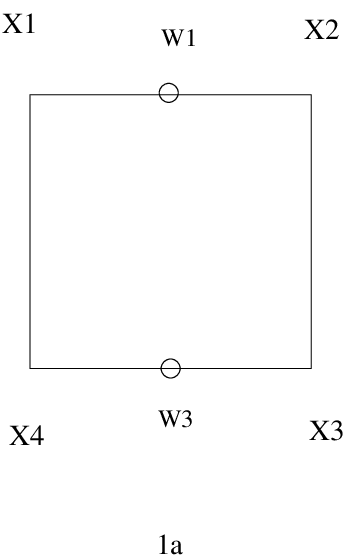}
  \quad
  \includegraphics[width=.315\textwidth]{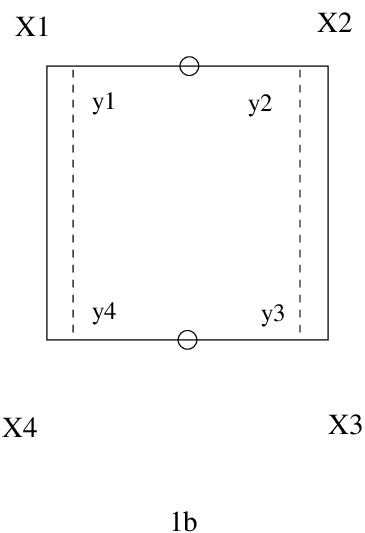}
  \caption{Projection on the $y^1-y^2$ plane of the 4-gluon scattering amplitude
 of \cite{Alday:2007hr} on the left
  and of the solution \eqref{Poincare} on the right.
  $X_1,X_2,X_3$ and $X_4$ are the positions of the cusps.
  The dashed lines denote
  the tips of the string which almost touch the boundary when the spin $S\rightarrow \infty$.
  The string  lies on the boundary from $y_1$ to $y_2$ at $\tau=-\infty$ and propagates
  in the bulk to end on the boundary again from $y_3$ to $y_4$ at $\tau=\infty$. $w_1=(y_2-y_1)/2$ and $w_3=(y_4-y_3)/2$
  are the points where the twist 2 operators
  ${\cal O}_S$ and ${\bar{\cal O}}_S$ are situated.
  Furthermore, $|w_{13}|=2 a$.}
  \label{fig:Figure1ab}
\end{figure}

Consequently, we have seen that the analytic continuation of the large spin, $S/\sqrt{\lambda}\rightarrow \infty$,
string solution \eqref{GKP} produces the square Wilson loop of \cite{Alday:2007hr}
which describes the scattering of four gluons at strong coupling.

In what follows, we will 
evaluate the 2-point correlator of
two twist 2 operators at strong coupling.
To this end, we need to calculate the action of the Euclidean world-sheet
solution of \eqref{GKP3}. 
It will be more convenient to rewrite
\eqref{GKP3} in the Poincare patch\footnote{To pass from the embedding coordinates
to the Poincare ones we have used the relations $Y^{\mu}=\frac{y^{\mu}}{z}$,
where $\mu=0,1,2,4$ with the direction 0 playing the role of time and $Y^{-1}+Y^{3}=\frac{1}{z}$,
$Y^{-1}-Y^{3}=\frac{z^2+y^{\mu}y_{\mu}}{z^2}$. }. It reads:
\begin{eqnarray}\label{Poincare}
y^1=a\tanh{\rho}\frac{\cosh{\phi}}{\cosh{t}}, \,\,\, y^4=a\tanh{t},\,\,\,
y^0=a\tanh{\rho}\frac{\sinh{\phi}}{\cosh{t}}, \,\,\,z=\frac{a}{\cosh{t}\cosh{\rho}},\,\,\\ \nonumber
\rho=\rho(\sigma),\, \phi=w \tau, \,t= k \tau
\end{eqnarray}
where $a$ is an overall scale which we have introduced by rescaling
all the Poincare coordinates.
This is a solution of the equations of motion in conformal gauge,
whose action is
\begin{eqnarray}\label{action}
S_E=\frac{R^2}{4 \pi \alpha'}\int d\sigma d\tau
\frac{\partial_{a}z\partial^{a}z+\partial_{a}y_{\mu}\partial^{a}y^{\mu}}{z^{2}},
\end{eqnarray}
where $a=1,2$ is a worldsheet index which is lowered-raised by $\eta_{ab}=\delta_{ab}$,
while $\mu$ is a spacetime index which is lowered-raised by the flat space time metric.
The isometry of the metric related to the $AdS_5$ angle $\phi$ leads to
the conserved spin ${\hat S}$ whose value is given by
\begin{eqnarray}\label{spin}
{\hat S}=\int d\sigma\frac{\partial {\cal L}_E}{\partial {\dot \phi}}=\frac{2 R^2}{ \pi \alpha'}\int_{0}^{\pi/2}d\sigma
\frac{y^{0}\partial_{\tau}y^{1}-\partial_{\tau}y^{0}y^{1}}{z^{2}}=
-\frac{2 R^2}{ \pi \alpha'}\int_{0}^{\pi/2}d\sigma \sinh^2{\rho}\,w=-S
\end{eqnarray}

As in the case of the four gluon scattering amplitude \cite{Alday:2007hr}
the action \eqref{action} is infinite when calculated on
the solution \eqref{Poincare}. We will choose to regulate the infinities by considering
the same solution on the background
\begin{eqnarray}\label{back}
ds^2=\frac{\sqrt{\lambda_D c_D}}{z^{\epsilon}}\big(\frac{dy_{4-2 \epsilon}^2+dz^2}{z^2} +d\Omega^2_{5+2 \epsilon}\big).
\end{eqnarray}
This background is obtained by performing a series of  T-dualities along the $D=4- 2\epsilon$ directions of the
string background describing the near-horizon limit of a stack of $D3-2 \epsilon$ branes \cite{Alday:2007hr}.

The dimensionally regulated action
is:
\begin{eqnarray}\label{reg}
S_E=\frac{\sqrt{\lambda_D c_D}}{4 \pi }\int d\sigma d\tau \frac{{\cal L}_{\epsilon=0}}{z^{\epsilon}},
\end{eqnarray}
where ${\cal L}_{\epsilon=0}$ is the Lagrangian density of
\eqref{action}. Furthermore, $\lambda_D=\frac{\lambda \mu^{2 \epsilon}}{(4 \pi e^{-\gamma})^{\epsilon}}$
and $c_D=2^{4 \epsilon} \pi^{3 \epsilon}\Gamma(2+\epsilon)$.
An important comment is in order. In general the exact solution
to the equations of motion derived from \eqref{reg} will depend on $\epsilon$
and is not known.
However, we need to compute the action
up to terms of order $O(\epsilon^0)$. For this it is enough to know the solution
up to order $O(\epsilon^0)$ too, that is the unperturbed solution.
The reason for this is that the action is divergent like $1/\epsilon$
\footnote{This is so because the $\sigma $ integration in the action is finite and the simple pole in $\epsilon$ comes from the $\tau$ integration.}.
Then apparently one needs to know the solution up to order $O(\epsilon)$.
But this is not necessary because when expanding the Lagrangian density of the action
${\cal L}(\phi)={\cal L}_0(\phi_0+\epsilon \phi_1)+ \epsilon {\cal L}_1(\phi_0+\epsilon \phi_1)=
({\cal L}_0+\epsilon {\cal L}_1)(\phi_0)+ \epsilon \frac{\delta {\cal L}_0(\phi)}{\delta\phi}|_{\phi=\phi_0} \phi_1+O(\epsilon^2)$ in terms of $\epsilon$, the
second term $\epsilon \frac{\delta {\cal L}_0(\phi)}{\delta\phi}|_{\phi=\phi_0} \phi_1$ is zero due to the equations of motion derived from ${\cal L}_0$. I
n the relations above we have used the expansions ${\cal L}={\cal L}_0+\epsilon {\cal L}_1+...$ and $\phi=\phi_0+\epsilon \phi_1+...$, where $\phi$
is an arbitrary field of the Lagrangian and $\phi_0$ the solution to the equations of motion derived from $L_0$.
Consequently, to evaluate the action up to order $\epsilon^0$ we need the the zeroth order solution, which we know, and
the dimensionally regularized Lagrangian which we also know.

We are now in position to evaluate the action $S_E^{(sol)}$ of the solution \eqref{Poincare}
that corresponds to a string propagating from boundary to boundary.
It is given by
\begin{eqnarray}\label{actsol}
S_E^{(sol)}=\int_{-\infty}^{\infty}d\tau\int_{0}^{\pi/2}d\sigma\frac{\sqrt{\lambda_D c_D}}{\pi a^{\epsilon}}
\cosh^\epsilon{k \tau}\cosh^\epsilon{\rho}(\rho'^2+ k^2 \cosh^2{\rho}-w^2\sinh^2{\rho}) =\nonumber \\
\int_{-\infty}^{\infty}d\tau\cosh^\epsilon{k \tau}\int_{0}^{\pi/2}d\sigma\frac{\sqrt{\lambda_D c_D}}{\pi a^{\epsilon}}
\cosh^\epsilon{\rho}\,2\,( k^2 \cosh^2{\rho}-w^2\sinh^2{\rho}).
\end{eqnarray}
To get the second line of \eqref{actsol} we have used the Virasoro constraint \eqref{rho'}.
One final step is needed before we are in position to
calculate the 2-point function. Namely, we should perform a Legendre
transformation of the action with respect to the angle $\phi$,
in the spirit of \cite{Janik:2010gc,Costa:2010rz,Tsuji:2006zn}.
\begin{eqnarray}\label{Legendre}
{\tilde S}_E=S_E^{(sol)}-\int_{-\infty}^{\infty} d\tau \,{\hat S} \, {\dot \phi}.
\end{eqnarray}
The evaluation of the second term on the right hand side of \eqref{Legendre}.
reads:
\begin{eqnarray}\label{Legendre2}
\int_{-\infty}^{\infty} d\tau \,{\hat S} \, {\dot \phi}=\frac{2\sqrt{\lambda_D c_D}}{\pi}
\int_{-\infty}^{\infty} d\tau \,\int_{0}^{\pi/2} d\sigma \,
\frac{y^{0}\partial_{\tau}y^{1}-\partial_{\tau}y^{0}y^{1}}{z^{2+\epsilon}}\, {\dot \phi}\nonumber \\
=-\frac{2\sqrt{\lambda_D c_D}}{ a^\epsilon \pi}\int_{-\infty}^{\infty} d\tau \cosh^{\epsilon}{k \tau}
\int_{0}^{\pi/2} d\sigma \,\sinh^2{\sigma}\cosh^{\epsilon}{\sigma} w^2.
\end{eqnarray}
Substituting \eqref{Legendre2} and \eqref{actsol} in \eqref{Legendre} one gets
\begin{eqnarray}\label{Legendre3}
{\tilde S}_E=\int_{-\infty}^{\infty}d\tau\cosh^\epsilon{k \tau}\int_{0}^{\pi/2}d\sigma\frac{\sqrt{\lambda_D c_D}}{\pi a^{\epsilon}}
\cosh^\epsilon{\rho}\,2\, k^2 \cosh^2{\rho}
\end{eqnarray}
It is now possible to perform the $\tau$ integration to get
\begin{eqnarray}\label{tau}
\frac{\sqrt{\lambda_D c_D}}{ a^{\epsilon}}\int_{-\infty}^{\infty}d\tau\cosh^\epsilon{k \tau}=
\frac{\sqrt{\lambda_D c_D}}{ a^{\epsilon}k}\frac{\sqrt{\pi}\Gamma(-\epsilon/2)}{\Gamma(1/2-\epsilon/2)}=
-\frac{2 \sqrt{\lambda}}{k \epsilon}-\frac{\sqrt{\lambda}}{k}(1+ \log{4 \pi^2}+\log{\frac{\mu^2}{w^2_{13}}}),\nonumber\\
\end{eqnarray}
where we used $w^2_{13}= 4 a^2$ to express the scale $a$ appearing in the solution in terms of the distance between
the operator insertions (see Figure 1b).
At the same time the $\rho$ integration gives
\begin{eqnarray}\label{rho}
\int_{0}^{\pi/2}d\sigma \frac{2}{\pi}\cosh^\epsilon{\rho} \cosh^2{\rho}\, k^2=
\frac{2}{\pi}\Big(\int_{0}^{\pi/2}d\sigma  k^2 \cosh^2{\rho}+
\epsilon Y \Big)=
k \frac{E}{\sqrt{\lambda}}+\frac{2}{\pi} \epsilon Y,
\end{eqnarray}
where Y is the integral originating from the expansion of $\cosh^\epsilon{\rho}=1+\epsilon \log{\cosh{\rho}}+O(\epsilon^2)$.
\begin{eqnarray}\label{why}
Y=\int_{0}^{\pi/2}d\sigma \log{\cosh{\rho}}\, k^2 \cosh^2{\rho}
\end{eqnarray}
It has a finite value which we will not need in what follows.
We should also mention that the value of the energy E  appearing in the last equality
of \eqref{rho} are the exact ones. There is no large $S/\sqrt{\lambda}>>1$ approximation.
Plugging \eqref{tau} and \eqref{rho} in the expression for the action \eqref{Legendre3} we get
\begin{eqnarray}\label{actsolfin}
\tilde{S}_E=-\frac{2 \sqrt{\lambda}}{\epsilon}\frac{E}{\sqrt{\lambda}}
-4\frac{\sqrt{\lambda}Y}{k \pi}-\sqrt{\lambda}(1+\log{4 \pi^2})\frac{E}{\sqrt{\lambda}} -\sqrt{\lambda}\frac{E}{\sqrt{\lambda}}\log{\frac{\mu^2}{w^2_{13}}}
\end{eqnarray}

Having obtained the Euclidean action \eqref{Legendre3} we can now
make use of the AdS/CFT correspondence to write down the
2-point function of two twist 2 operators, at strong coupling, as
\begin{eqnarray}\label{2point}
<{\bar {\cal O}}_S(w_3)\,\,{\cal O}_S(w_1)>=e^{-{\tilde S}_E}=e^{\frac{2}{\epsilon}E
+4\frac{\sqrt{\lambda}Y}{k \pi}+(1+\log{4 \pi^2})E}\,\Big( \frac{\mu^2}{w_{13}^2}\Big)^{\Delta_S},
\end{eqnarray}
where $\Delta_S=E$ is the exact
scaling dimension of a twist 2 operators with arbitrary spin $S$.
We stress once more that the strong coupling 2-point function \eqref{2point} is valid for any value of the spin $S/\sqrt{\lambda}$
and is not restricted to
twist two operators of large spin since in the string solution used to obtain \eqref{2point} no large $S$, i. e. $S/\sqrt{\lambda}>>1$,
approximation has been made.

Before ending this Section we would like to comment on the spacetime dependence
of \eqref{2point}. 
We have chosen to put the two operators at $w_1=(y_2-y_1)/2$ and $w_3=(y_4-y_3)/2$, respectively.
This association is, somehow, ambiguous from the sting theory point of view
because the tunneling string is extending from $y_1$ to $y_2$ at $\tau=-\infty$
and similarly from $y_3$ to $y_4$ at $\tau=\infty$.
However, it seems natural to place the operators at the center of mass of the string when this is touching the boundary, that is at $w_1$ and $w_3$.


Another way to justify the insertion of the operators at the points
$w_1$ and $w_3$ is the following \cite{Alday:2010ku}. Imagine starting by putting two operators
${\cal O}_S(w_{1})=
\tr{\phi_1(w_{1})\overleftarrow{D}_+^{S/2}\overrightarrow{D}_+^{S/2}\phi_2(w_{1})}$  and
$\bar{{\cal O}}_S(w_{3})=\tr{\phi_2(w_{3})\overleftarrow{D}_-^{S/2}\overrightarrow{D}_-^{S/2}\phi_1(w_{3})}$ at a space-like distance $w_{13}^2=4 \,a^2$. The presence of a number of covariant derivatives
is effectively producing  a displacement of the scalar fields along the light-cone direction of the derivative.
As the spin grows, the effective displacement also grows, and in the limit of infinite spin
the displaced insertions of the first operator, which approach $x_1$ and $x_3$, become light-like separated with respect to the
displaced scalar fields of the second operator, which approach $x_3$ and $x_4$.
The displaced scalar fields of each operator are, of course, joined by a Wilson line.
So, what we really need to evaluate 
is the correlator $<W_1[y_1,y_2] \,W_2[y_3,y_4]>$
of two Wilson lines $W[y_1,y_2]=Tr \big(P \phi_1(y_1)e^{i\int_{y_1}^{y_2}A_{\mu}dy^{\mu}}\phi_2(y_2)\big)$
and $W_2[y_3,y_4]=Tr \big(P \phi_2(y_3)e^{i\int_{y_3}^{y_4}A_{\mu}dy^{\mu}}\phi_1(y_4)\big)$
extending from  $y_1$ to  $y_2$ and from  $y_3$ to $y_4$, respectively (see Figure 1b).
In the limit of infinite spin a four-sided Wilson loop emerges.

Finally, let us mention that one can put \eqref{2point} to the canonical form
by defining renormalised operators through the relation
\begin{equation}\label{renormalised}
{\cal O}_s^{ren.}={\cal Z}^{-\frac{1}{2}}{\cal O}_s,
\end{equation}
where
\begin{equation}\label{Z}
{\cal Z}=e^{\frac{2}{\epsilon}E
+4\frac{\sqrt{\lambda}Y}{k \pi}+(1+\log{(4 \pi^2 \mu^2)})E}.
\end{equation}
Then the 2 point function of the renormalised operators will take the form
\begin{equation}\label{2pointren}
<{\bar {\cal O}}_S^{ren.}(w_3)\,\,{\cal O}_S^{ren.}(w_1)>=\frac{1}{w^{2\Delta_S}_{13}}.
\end{equation}
\subsection{2-point function of twist $J$ operators}\label{twistJ}

We now turn to the case of a large spin operator which also carries a large
angular momentum $J=\sqrt{\lambda} {\cal J}$ around the equator of $S^5$.
The operator can be written schematically as
\begin{equation}\label{twistJop}
{\cal O}_{SJ}=\tr{D^S_+Z^J}+...
\end{equation}
According to AdS/CFT there should exist a semiclassical string solution that
corresponds to this operator. This solution was found in \cite{Frolov:2002av}
and in global coordinates is
\begin{eqnarray}\label{GKPJ}
t=k \tau, \, \phi=w t, \, \rho=\rho(\sigma), \, \phi_1=\nu \tau, \,\rho \in [0,\rho_0],\\
E_{SJ}=\frac{2\sqrt{\lambda}}{\pi}\,k \int_{0}^{\pi/2} d\sigma \cosh^2{\rho},\, J=\sqrt{\lambda} \nu ,\, S=\frac{2\sqrt{ \lambda}}{\pi}\, w\,\int_{0}^{\pi/2} d\sigma \sinh^2{\rho},
\end{eqnarray}
where $\phi_1$ parametrises a maximal circle of the $S^5$,
$w'^2=\frac{w^2-\nu^2}{k^2-\nu^2}=\cosh^2{\rho_0}/\sinh^2{\rho_0}$ and $\rho(\sigma)$ is the solution of the
equation
\begin{equation}\label{rho'J}
\rho'^2=k^2
\cosh^2{\rho}-w^2\sinh^2{\rho}-\nu^2.
\end{equation}

In the large $S$ limit its energy is given by
\begin{equation}\label{GKPJenergy}
E=S+\frac{\sqrt{\lambda}}{\pi} k \log{\frac{S}{\sqrt{\lambda}}}
\end{equation}
but we will not use this approximation and will keep the value of $S$ arbitrary.
One can now perform the analytic continuation of \eqref{GKPJ} according to \eqref{continuation}. We also choose to
leave the $S^5$ angle untouched $\phi_1 \rightarrow  \phi_1$.
The result of this is to get the real solution
\begin{eqnarray}\label{GKPJanalytic}
Y^{-1}=\cosh{t}\cosh{\rho},  \qquad Y^0=\sinh{\phi} \sinh{\rho},
\qquad Y^1=\cosh{\phi}\sinh{\rho}, \nonumber \\
\qquad Y^4=\sinh{t}\cosh{\rho}, \qquad Y^2=Y^3=0,
\qquad \rho=\rho(\sigma),\,t= k \tau ,\, \phi=w \tau, \,\phi_1= i \nu \tau.
\end{eqnarray}
which can also be written in Poincare coordinates as
\begin{eqnarray}\label{PoincareJ}
y^1=a\tanh{\rho}\frac{\cosh{\phi}}{\cosh{t}}, \,\,\, y^4=a\tanh{t},\,\,\,
y^0=a\tanh{\rho}\frac{\sinh{\phi}}{\cosh{t}}, \,\,\,z=\frac{a}{\cosh{t}\cosh{\rho}},\,\,\\ \nonumber
\rho=\rho(\sigma),\, \phi=w \tau, \,t= k \tau,\,\phi_1=i \nu \tau
\end{eqnarray}
where as in \eqref{Poincare} $a$ is an overall scale which we have introduced by rescaling
all the Poincare coordinates.
As in the previous Section, we work in dimensional regularisation. Starting from the action
\begin{eqnarray}\label{actionSJ}
S_E=\frac{\sqrt{\lambda_D c_D}}{4 \pi }\int d\sigma d\tau \frac{{\cal L}_{\epsilon=0}+{\dot\phi}^2_1}{z^{\epsilon}},
\end{eqnarray}
where ${\cal L}_{\epsilon=0}$ is the Lagrangian density of
\eqref{action}, we perform a Legendre transform with respect to the two angles
that are conjugate to the two angular momenta $S$ and $J$ to get
\begin{eqnarray}\label{LegendreSJ}
{\tilde S}_E=S_E-\int_{-\infty}^{\infty}d\tau J {\dot\phi}_1-\int_{-\infty}^{\infty}d\tau {\hat S} {\dot\phi}=
\frac{\sqrt{\lambda_D c_D}}{ \pi}\int_{0}^{\pi/2} d\sigma d\tau
\frac{{\cal L}_{\epsilon=0}-{\dot\phi}^2_1}{z^{\epsilon}}-\int_{-\infty}^{\infty}d\tau {\hat S} {\dot\phi}.
\end{eqnarray}
Substituting the solution \eqref{PoincareJ} in \eqref{LegendreSJ} we get
\begin{eqnarray}\label{LegendreSJ1}
{\tilde S}_E=\int_{-\infty}^{\infty}d\tau\int_{0}^{\pi/2}d\sigma\frac{\sqrt{\lambda_D c_D}}{\pi a^{\epsilon}}
\cosh^\epsilon{k \tau}\cosh^\epsilon{\rho}(\rho'^2+ k^2 \cosh^2{\rho}-w^2\sinh^2{\rho}+\nu^2)\nonumber \\-\int_{-\infty}^{\infty}d\tau {\hat S} {\dot\phi} =
\int_{-\infty}^{\infty}d\tau\cosh^\epsilon{k \tau}\int_{0}^{\pi/2}d\sigma\frac{\sqrt{\lambda_D c_D}}{\pi a^{\epsilon}}
\cosh^\epsilon{\rho}\,2\,( k^2 \cosh^2{\rho}-w^2\sinh^2{\rho})\nonumber\\
-\int_{-\infty}^{\infty}d\tau {\hat S} {\dot\phi}=\int_{-\infty}^{\infty}d\tau\cosh^\epsilon{k \tau}\int_{0}^{\pi/2}d\sigma\frac{\sqrt{\lambda_D c_D}}{\pi a^{\epsilon}}
\cosh^\epsilon{\rho}\,2\, k^2 \cosh^2{\rho}.
\end{eqnarray}
As in the previous Subsection we have used the Virasoro constraint \eqref{rho'J} to pass from the first to the
second line of \eqref{LegendreSJ1}.
Then one can precisely follow the steps leading from \eqref{Legendre3} to \eqref{2point} and \eqref{2pointren} to get
\begin{equation}\label{2pointrenJ}
<{\bar {\cal O}}_{SJ}^{ren.}(w_3)\,\,{\cal O}_{SJ}^{ren.}(w_1)>=\frac{1}{(w^2_{13})^{\Delta_{SJ}}},
\end{equation}
where $\Delta_{SJ}=E_{SJ}$ is the exact in $S$
scaling dimension of the operators in \eqref{twistJop}.

\subsection{2-point function of operators with a large spin in $AdS_5$ and two equal spins in $S^5$}\label{twistJJ}

Finally, let us consider the case of string that has a large spin in  $AdS_5$ but is also spinning
in $S^5$ with two equal angular momenta $J_1=J_2=J/2$. More precisely, the $AdS$ part of the solution
will be similar to the solutions consider so far.
Namely, it will be a folded string extending around the centre of $AdS_5$.
In $S^5$, we will have a circular string that is winding $n$-times around one of the non-isometric angles of the sphere \footnote{Here we parametrise the five-sphere by the angles $(\theta,a,\phi_1,\phi_2,\phi_3)$,\\
${\bold n}=n_i=(\sin{\theta}\cos{\phi_1},\sin{\theta}\sin{\phi_1},\cos{\theta}\sin{a}\cos{\phi_2},
\cos{\theta}\sin{a}\sin{\phi_2},\cos{\theta}\cos{a}\cos{\phi_3},\cos{\theta}\cos{a}\sin{\phi_3})$.},
$\theta$, but is also rotating with equal angular momenta around two of the isometries $\phi_1$ and $\phi_2$.
Although this solution is similar to the one considered in the previous Subsection, three point correlators involving
it will be drastically different compared to those involving the solution of Section \ref{twistJ}.

Firstly, let us discuss the structure of this solution in some detail.
Then we will analytically  continue the solution according to \eqref{continuation}.
Lastly, we will perform the Legendre transform of the analytically  continued solution
to obtain the two-point function at strong coupling.

The string solution we are seeking will satisfy the equations of motion derived from the Polyakov action,
as well as, the Virasoro constraints. By considering the following ansatz
\begin{eqnarray}\label{ansatzJJ}
t=k \tau,\qquad \phi=w \tau, \qquad \rho=\rho(\sigma),\nonumber\\
a=\frac{\pi}{2} ,\qquad\theta=\theta(\sigma),\qquad \phi_1=\nu_1 \tau,\qquad \phi_2=\nu_2 \tau,\qquad \nu_1=\nu_2
\end{eqnarray}
we can write the action in the form
\begin{equation}\label{PactionJJ}
S=S_{AdS_5}+\frac{\sqrt{\lambda}}{4 \pi}\int d\tau \int_{0}^{2 \pi} d\sigma
(\theta'^2-\cos^2{\theta} {\dot \phi}_2^2-\sin^2{\theta} {\dot \phi}_1^2),
\end{equation}
where $S_{AdS_5}$ is the $AdS$ part of the action.
The equation of motion for $\theta$ is
\begin{equation}\label{psi}
\theta''+\sin{\theta}\cos{\theta}(\nu_1^2-\nu_2^2)=0
\end{equation}
which can be easily solved due to the condition $\nu_1=\nu_2$ to give
\begin{equation}\label{psi1}
\theta=n \sigma,\, n \in N^*, \,\sigma \in [0,2\pi].
\end{equation}
The meaning of \eqref{psi1} is, of course, that the string is wrapping the circle parametrised by
$\theta$ $n$ times.
The equations of motion for the coordinates of the $AdS_5$ space are the same as those of the solution
in Section \ref{twistJ}. The only difference will come through the non-trivial Virasoro constraint which
will become
\begin{equation}\label{Virasoro}
\rho'^2=k^2 \cosh^2{\rho}- w^2 \sinh^2{\rho}-{\tilde \nu}^2,\qquad {\tilde \nu}^2=\nu^2+n^2.
\end{equation}
Notice that the only change with respect to the case of a single angular momentum (Section \ref{twistJ}) is
the substitution of $\nu$ by ${\tilde \nu}$. Given this one can use the results of \cite{Frolov:2002av}
and immediately write down our solution in a form similar to that of Section \ref{twistJ}.
\begin{eqnarray}\label{solJJ}
t=k \tau, \, \phi=w \tau, \, \rho=\rho(\sigma), \,  \,
\theta=n \sigma,\, \phi_1=\phi_2=\nu \tau,\, \rho \in [0,\rho_0]\nonumber \\
E_{SJJ}=\frac{2\sqrt{\lambda}}{\pi}\int_{0}^{\pi/2} d\sigma \cosh^2{\rho}\,k, \,\,\, J=J_1+J_2=\sqrt{\lambda} \nu ,\, \,\, S=\frac{2\sqrt{ \lambda}}{\pi}\int_{0}^{\pi/2} d\sigma \sinh^2{\rho}\,w,
\end{eqnarray}
where $\frac{w^2-\nu^2-n^2}{k^2-\nu^2-n^2}=\cosh^2{\rho_0}/\sinh^2{\rho_0}$ and $\rho(\sigma)$ is the solution of the
\eqref{Virasoro}.

The analytic continuation of \eqref{solJJ}, according to \eqref{continuation},
written in Poincare coordinates is
\begin{eqnarray}\label{PoincareJJ}
y^1=a\tanh{\rho}\frac{\cosh{\phi}}{\cosh{t}}, \,\,\, y^4=a\tanh{t},\,\,\,
y^0=a\tanh{\rho}\frac{\sinh{\phi}}{\cosh{t}}, \,\,\,z=\frac{a}{\cosh{t}\cosh{\rho}},\,\,\\ \nonumber
\rho=\rho(\sigma),\, \phi=w \tau, \,t= k \tau,\,\theta=n \sigma ,\,\,\phi_1=i \nu \tau=\phi_2
\end{eqnarray}
The corresponding action reads
\begin{eqnarray}\label{PactionSJJ}
S_E=\frac{\sqrt{\lambda_D c_D}}{4 \pi }\int d\sigma d\tau \frac{{\cal L}_{\epsilon=0}+(\theta'^2+\cos^2{\theta} {\dot \phi}_2^2+\sin^2{\theta} {\dot \phi}_1^2)}{z^{\epsilon}}.
\end{eqnarray}
The difference in sign between the ${\dot \phi}_1^2$ and ${\dot \phi}_2^2$ terms of \eqref{PactionSJJ}
and the same terms of \eqref{PactionJJ} is due to the Euclidean nature of the worldsheet in \eqref{PactionSJJ}
compared to the Lorentzian worldsheet of \eqref{PactionJJ}.

The next step is to perform a Legendre transform with respect to the angles
that are conjugate to the three angular momenta $S$ and $J_1$ and $J_2$ to get
\begin{eqnarray}\label{LegendreSJJ}
{\tilde S}_E=S_E-\int_{-\infty}^{\infty}d\tau (J_1 {\dot\phi}_1+J_2 {\dot\phi}_2)-
\int_{-\infty}^{\infty}d\tau {\hat S} {\dot\phi}=\nonumber\\
\frac{\sqrt{\lambda_D c_D}}{ \pi}\int_{0}^{\pi/2} d\sigma d\tau
\frac{{\cal L}_{\epsilon=0}+\theta'^2-\cos^2{\theta}{\dot\phi}^2_2-\sin^2{\theta}{\dot\phi}^2_1}{z^{\epsilon}}
-\int_{-\infty}^{\infty}d\tau {\hat S} {\dot\phi}=\nonumber\\\int_{-\infty}^{\infty}d\tau\cosh^\epsilon{k \tau}\int_{0}^{\pi/2}d\sigma\frac{\sqrt{\lambda_D c_D}}{\pi a^{\epsilon}}
\cosh^\epsilon{\rho}\,2\, k^2 \cosh^2{\rho}.
\end{eqnarray}
In deriving the last line of \eqref{LegendreSJJ} we have used as usual the Virasoro constraint \eqref{Virasoro}.

Following the steps leading from \eqref{Legendre3} to \eqref{2point} and \eqref{2pointren} we can
write the two point correlator of the operator dual to the
classical solution \eqref{solJJ} at strong coupling as
\begin{equation}\label{2pointrenJ}
<{\bar {\cal O}}_{SJJ}^{ren.}(w_3)\,\,{\cal O}_{SJ}^{ren.}(w_1)>=\frac{1}{(w^2_{13})^{\Delta_{SJJ}}},
\end{equation}
where $\Delta_{SJJ}=E_{SJJ}$ is the exact in $S$
scaling dimension of the operators dual to the string solution of \eqref{solJJ}
\footnote {Let us note that it would be interesting to evaluate by using the same technique the strong coupling 2-point correlator of  
classical string solutions \cite{Chu:2006ae} in $\beta$-deformed theories.}.

In the next Section, we will see that despite the fact that the two solutions \eqref{solJJ}
and \eqref{GKPJ} are very similar, the fusion coefficients involving their dual operators have very different
$S$ dependence.

\sect{3-point function of  two twist 2 operators and a BPS operator}
\label{sec:3-point}

In this Section, we will use the solutions obtained in the Section 2 to
calculate the 3 point function coefficient of two large spin operators and a
supergravity state, at strong coupling.
For simplicity, we will choose the BPS state to be the BMN vacuum
\begin{eqnarray}\label{CPO}
{\cal O}_I(x)\sim \tr {Z^l}(x),
\end{eqnarray}
where $Z=\Phi_1+i \Phi_2$ is one of the complex scalar fields of $N=4$ SYM.
Following the normalisations of \cite{Zarembo:2010rr} the
corresponding spherical harmonic is
\begin{eqnarray}\label{harmonic}
Y_I({\bold n})=\Big( \frac{n_1+n_2}{\sqrt{2}}\Big)^l=\frac{1}{2^{\frac{l}{2}}} \sin^l{\theta} e^{i l \phi_1},
\end{eqnarray}
where ${\bold n}$ is a six-dimensional vector defining a point on the 5-sphere,
\begin{eqnarray}\label{5sphere}
{\bold n}=n_i=(\sin{\theta}\cos{\phi_1},\sin{\theta}\sin{\phi_1},\cos{\theta}\sin{a}\cos{\phi_2},
\cos{\theta}\sin{a}\sin{\phi_2},\nonumber\\\cos{\theta}\cos{a}\cos{\phi_3},
\cos{\theta}\cos{a}\sin{\phi_3}).
\end{eqnarray}
In \eqref{5sphere} $\phi_1,\,\phi_2,\,\phi_3$ parametrise the three isometries of the sphere.
For all the solutions considered in this note $a=\pi/2$ which means that $n_5=n_6=0$.

The correlators on which we will focus are
\begin{eqnarray}\label{corr1}
<\bar{{\cal O}}_{SJ}(w_3)\,{\cal O}_{SJ}(w_1)\, {\cal O}_I(x)>
\end{eqnarray}
and
\begin{eqnarray}\label{corr2}
<\bar{{\cal O}}_{SJJ}(w_3)\,{\cal O}_{SJJ}(w_1)\, {\cal O}_I(x)>,
\end{eqnarray}
where ${\cal O}_{SJ}$ and ${\cal O}_{SJJ}$ are the operators dual to the string solutions
of Sections 2.2 and 2.3, respectively.
Before we evaluate the structure constants of the above correlators let us
quickly review the formalism of \cite{Zarembo:2010rr} which we will use in what follows.

\subsection{General formalism}
In order to calculate the 3-point function of a chiral
primary operator with two operators that correspond to semiclassical states
we follow \cite{Zarembo:2010rr,Costa:2010rz}.
In this approach one treats the BPS state in the supergravity approximation
while the massive string state is being treated in the first quantised string theory
by summing over all classical trajectories.
The quantity we would like to consider is \cite{Berenstein:1998ij,Zarembo:2010rr}
\begin{eqnarray}\label{Zar-1}
<{\cal O}_I(x)>_{{\bar{\cal O}}_J {\cal O}_J}=\frac{<{\bar{\cal O}}_J(w_3)\,{\cal O}_J(w_1)\,{\cal O}_I(x)>}
{<{\bar{\cal O}}_J(w_3)\,{\cal O}_J(w_1)>}.
\end{eqnarray}
In \eqref{Zar-1}, ${\cal O}_I$ is a  chiral primary operator while ${\cal O}_J$ is a primary
operator corresponding to a massive string state. By taking $|x|\rightarrow \infty$ and by using
\eqref{3pointconformal} we get
\begin{eqnarray}\label{Zar0}
<{\cal O}_I(x)>_{{\bar{\cal O}}_J {\cal O}_J}=\frac{C_{{\bar{\cal O}}_J {\cal O}_J{\cal O}_I}
\,|w_{13}|^{\Delta_I}}{x^{2\Delta_I}}
\end{eqnarray}
The strong coupling dual
of this type of correlator is \cite{Zarembo:2010rr}:
\begin{eqnarray}\label{Zar1}
<{\cal O}_I(x)>_{{\bar{\cal O}}_J {\cal O}_J}=\lim_{\epsilon \rightarrow 0} \frac{\pi}{\epsilon^{\Delta_I}}
\sqrt{\frac{2}{\Delta_I-1}}\frac{1}{\int D\Phi e^{-S_{sugra}[\Phi]}}
\int D\Phi \,\Phi_I(x,\epsilon) \frac{1}{Z_{str}} \int DX e^{-S_{str}[X,\Phi]} \nonumber \\,
\end{eqnarray}
where $\Phi$ are the supergravity fields and $X$ are the embedding coordinates of the
worldsheet in $AdS_5 \times S^5$. One important point about \eqref{Zar1} is the
dependence of the string action on the supergravity fields. This dependence enters the
string action indirectly through the perturbations of the metric, of the Kalb-Ramond field, of the dilaton,
and of the fermions created by the BPS operator insertion.
The way to proceed is clear.
One should expand the string action with respect to the supergravity fields  keeping terms
up to linear order in $\Phi_I$ to get
\begin{eqnarray}\label{Zar2}
<{\cal O}_I(x)>_{{\bar{\cal O}}_J {\cal O}_J}=-\lim_{\epsilon \rightarrow 0} \frac{\pi}{\epsilon^{\Delta_I}}
\sqrt{\frac{2}{\Delta_I-1}}
\frac{1}{Z_{str}} \int DX e^{-S_{str}[X,\Phi=0]}
\frac{1}{\int D\Phi e^{-S_{sugra}[\Phi]}}\nonumber\\
\int d^2\sigma \int D\Phi\, \Phi_I(x,\epsilon)
\frac{\delta S_{str}[X,\Phi=0]}{\delta \Phi_I(y(\sigma,\tau),z(\sigma,\tau))} \Phi_I(y,z)
\end{eqnarray}
Then the $\Phi$ path integral can be performed giving, essentially, a second order differential operator
with $X$ dependent coefficients acting on the BPS state's bulk to boundary propagator \cite{Zarembo:2010rr}.
Finally, the $X$ path integral will be, as usual, dominated by  the classical string solution
(in our case the solution of Section 2). The result has the form of a vertex operator integrated
over the classical worldsheet.

Substituting in \eqref{Zar2} the expressions for the aforementioned differential operator
and taking care of the normalisations of the sugra states one gets
for the OPE coefficient \cite{Zarembo:2010rr}
\begin{eqnarray}\label{CPOstrong}
C_{{\bar{\cal O}}_J {\cal O}_J{\cal O}_I} \,|w_{13}|^{\Delta_I}=\frac{2^{{\frac{l}{2}-3}}(l+1) \sqrt{l \lambda}}{\pi N}\int d\tau d\sigma
Y_I({\bold n})z^l(\frac{\partial_a X^{\mu}\partial^a X_{\mu}-\partial_a z\partial^a z}{z^2}-
\partial_a \bold{n}\partial^a \bold{n}).
\end{eqnarray}
For the full details, see \cite{Zarembo:2010rr}. Let us only mention that
$Y_I({\bold n})$ is the spherical harmonic corresponding to the chiral primary
participating in the 3-point correlator.

\subsection{Evaluation of the correlator $<\bar{{\cal O}}_{SJ}(w_3)\,{\cal O}_{SJ}(w_1)\, {\cal O}_I(x)>$}

We can now make use of \eqref{CPOstrong} in order to evaluate the fusion coefficient of \eqref{corr1}.
By plugging in \eqref{CPOstrong} the classical solution \eqref{PoincareJ}
which
corresponds to the non-BPS twist J operator of \eqref{twistJop} one gets
\begin{eqnarray}\label{3CPO}
C_{{\bar O}_{SJ}  O_{SJ} {\cal O}_I}=\frac{(l+1) \sqrt{l \lambda}}{2^l \pi N}\int_{-\infty}^{\infty} d\tau \int_{0}^{\pi/2}d\sigma
\frac{e^{-l \nu \tau}}{\cosh^l{k \tau}\cosh^l{ \rho}}(\frac{\rho'^2}{\cosh^2{\rho}}
-k^2\tanh^2{k \tau}+\nu^2)\nonumber\\.
\end{eqnarray}
Here we have also made use of the fact that $|w_{13}|=2 a$ and that the BPS operator has dimension $\Delta_I=l$.
The two first terms inside the integral come from the $AdS_5$ part of the solution while the $\nu^2$ term from
the $S^5$ part.
It may appear natural that if one is interested in long ($S\rightarrow \infty$) string solutions one could
substitute in \eqref{3CPO} the approximate solution $\rho= \mu \sigma,...$ and then perform the integrals to obtain
the 3-point coefficient. Indeed, if we substitute this solution in \eqref{3CPO} we get equation (4.25) of \cite{Roiban:2010fe} and as a consequence the same fusion coefficient.

However, this approximate large $S$ solution is not accurate enough to give the $1/S$ corrections to the fusion coefficient.
Here we choose to use the exact string solution when calculating the integrals in \eqref{3CPO}.
Doing so and rewriting the Virasoro constraint as
\begin{eqnarray}\label{Virasoronew}
\rho'=\sqrt{k^2-\nu^2}\sqrt{\cosh^2{\rho}-w'^2\sinh^2{\rho}},\,\,w'^2=\frac{w^2-\nu^2}{k^2-\nu^2}
\end{eqnarray}
we obtain from \eqref{3CPO}
\begin{eqnarray}\label{3CPO1}
C_{{\bar O}_{SJ}  O_{SJ} {\cal O}_I}=\frac{(l+1) \sqrt{l \lambda}}{2^l \pi N}\Big(\int_{-\infty}^{\infty} d\tau
\frac{e^{-l \nu \tau}}{\cosh^l{k \tau}}\int_{0}^{\rho_0}d\rho
\frac{\sqrt{k^2-\nu^2}\sqrt{\cosh^2{\rho}-w'^2\sinh^2{\rho}}}{\cosh^{l+2}{ \rho}}+\nonumber\\
\int_{-\infty}^{\infty} d\tau
\frac{e^{-l \nu \tau}(\nu^2-k^2\tanh^2{k\tau})}{\cosh^l{k \tau}}\int_{0}^{\rho_0}d\rho
\frac{1}{\sqrt{k^2-\nu^2}\sqrt{\cosh^2{\rho}-w'^2\sinh^2{\rho}}\cosh^l{ \rho}}
\Big)\nonumber\\.
\end{eqnarray}

We see that the integrals with respect to $\tau$ and $\sigma$ disentangle and can be performed
separately. The $\tau$ integrals that we need are of the form
\begin{eqnarray}\label{tauint1}
I_{\tau 1}=\int_{-\infty}^{\infty} d\tau
\frac{e^{-l \nu \tau}}{\cosh^l{k \tau}}=2^{l}\Big(\frac{\Gamma(\frac{1}{2}(l-\frac{l \nu}{k}))\Gamma(\frac{1}{2}(l+\frac{l \nu}{k}))}{k \Gamma(l)}
-\frac{{}_2F_1(\frac{1}{2}(l+\frac{l \nu}{k}),l;\frac{1}{2}(l+2+\frac{l \nu}{k});-1)}{l(k+ \nu)}\nonumber\\
-\frac{{}_2F_1(\frac{1}{2}(l-\frac{l \nu}{k}),l;\frac{1}{2}(l+2-\frac{l \nu}{k});-1)}{l(k-\nu)}
\Big).
\end{eqnarray}
and
\begin{eqnarray}\label{tauint2}
I_{\tau 2}=\int_{-\infty}^{\infty} d\tau
\frac{e^{-l \nu \tau}}{\cosh^{l+2}{k \tau}}=2^{l+2}\Big(\frac{\Gamma(\frac{1}{2}(l+2-\frac{l \nu}{k}))\Gamma(\frac{1}{2}(l+2+\frac{l \nu}{k}))}{k \Gamma(l+2)}\nonumber\\
-\frac{{}_2F_1(\frac{1}{2}(l+2+\frac{l \nu}{k}),l+2;\frac{1}{2}(l+4+\frac{l \nu}{k});-1)}{k(l+2)+l \nu}\nonumber\\
-\frac{{}_2F_1(\frac{1}{2}(l+2-\frac{l \nu}{k}),l+2;\frac{1}{2}(l+4-\frac{l \nu}{k});-1)}{k(l+2)-l \nu}
\Big).
\end{eqnarray}
We now turn to the $\sigma$ integrals that we need. These are obtained by performing the following change of variables
$w' \tanh{\rho}=\sin{\theta}$
\begin{eqnarray}\label{sigmaint1}
I_{\sigma 1}=\int_{0}^{\rho_0}d\rho
\frac{\sqrt{k^2-\nu^2}\sqrt{\cosh^2{\rho}-w'^2\sinh^2{\rho}}}{\cosh^{l+2}{ \rho}}=\int_{0}^{\pi/2}d\theta \frac{\sqrt{k^2-\nu^2}}{w'}\cos^2{\theta}\Big(\sqrt{1-\frac{\sin^2{\theta}}{w'^2}}\Big)^{l-1}\nonumber\\
=\frac{\sqrt{k^2-\nu^2}}{4 w'^3}(1-\frac{1}{w'^2})^{\frac{l-1}{2}}\pi\Big(3(w'^2-1)\,
{}_2F_1(-\frac{1}{2},\frac{1-l}{2};2;\frac{1}{1-w'^2})+\nonumber\\
(l+2-2w'^2)\,{}_2F_1(\frac{1}{2},\frac{1-l}{2};2;\frac{1}{1-w'^2}) \Big).\nonumber\\
\end{eqnarray}
Similarly, the other integral with respect to $\sigma$ gives
\begin{eqnarray}\label{sigmaint1}
I_{\sigma 2}=\int_{0}^{\rho_0}d\rho
\frac{1}{\sqrt{k^2-\nu^2}\sqrt{\cosh^2{\rho}-w'^2\sinh^2{\rho}}\cosh^{l}{ \rho}}=\nonumber\\
\int_{0}^{\pi/2}d\theta
\frac{1}{\sqrt{k^2-\nu^2}w'}\Big(\sqrt{1-\frac{\sin^2{\theta}}{w'^2}}\Big)^{l-1}
=\frac{1}{\sqrt{k^2-\nu^2}2 w'}\pi\,
{}_2F_1(\frac{1}{2},\frac{1-l}{2};2;\frac{1}{w'^2}).
\end{eqnarray}

We can now write down the fusion coefficient as
\begin{eqnarray}\label{3CPO2}
C_{{\bar O}_{SJ}  O_{SJ} {\cal O}_I}=\frac{(l+1) \sqrt{l \lambda}}{2^l \pi N}\Big(
I_{\tau 1}I_{\sigma 1}+I_{\sigma 2}\big((\nu^2-k^2)I_{\tau 1}+k^2 I_{\tau 2}\big)\Big)
\end{eqnarray}

\eqref{3CPO2} is the main result of this Section. It is valid for 3-point correlators involving
long, $S/\sqrt{\lambda}>>1$, or short, $S/\sqrt{\lambda}<<1$, string solutions with ($\nu\neq 0$) or without
($\nu=0$) angular momentum in $S^5$. One can, in principle, express all the parameters appearing in  the right hand side of \eqref{3CPO2}
in terms of the two angular momenta of the string $S$ and $J$.

It is possible take various limits of \eqref{3CPO2}.
For instance, one can expand \eqref{3CPO2} for small $\nu$, keep the leading term and then take the large spin limit $S/\sqrt{\lambda}>>1$.
This corresponds to the

$\bullet$ Long string solution

The resulting expression for the fusion coefficient becomes
\begin{eqnarray}\label{smallnuJ}
C_{{\bar O}_{SJ}  O_{SJ} {\cal O}_I}=\frac{(l+1)\, \sqrt{l \lambda}}{2^{l+2} N \Gamma(\frac{l+3}{2})}
\Big(\frac{(l-1)\Gamma^2(\frac{l}{2})}{\Gamma(\frac{l+1}{2})} -\frac{(l-2)\Gamma^2(\frac{l}{2}-1)}{2 \pi \Gamma(\frac{l-1}{2})} \frac{1}{S/\sqrt{\lambda}}\Big)\nonumber\\
+O(\frac{1}{(S/\sqrt{\lambda})^{\frac{3}{2}}})+O(\nu),\,for \, l>2\nonumber\\
C_{{\bar O}_{SJ}  O_{SJ} {\cal O}_I}=\frac{ \sqrt{\lambda}}{2^{1/2} \pi N}-\frac{\sqrt{2\lambda}}{4\pi^2 N}(\log{8}-1)
\frac{1}{S/\sqrt{\lambda}}\nonumber\\
+O(\frac{1}{(S/\sqrt{\lambda})^{\frac{3}{2}}})+O(\nu),\,for \, l=2
\end{eqnarray}

In \eqref{smallnuJ} we have expanded the hypergeometric functions around $w'=1$. The relevant expansions of
$I_{\sigma 1},I_{\sigma 2},I_{\tau 1},I_{\tau 2}$ are given in the Appendix.
The first correction (subleading term) is proportional to
$w'-1\approx \frac{1}{\pi}\frac{1}{S/\sqrt{\lambda}}$.
We see that the leading term of the expansion is a constant that depends only on the BPS state and not on
the quantum numbers of the massive string state.

As mentioned above, only the leading term of \eqref{smallnuJ} is in agreement  with
the corresponding  equation of \cite{Roiban:2010fe} (4.28),
where the fusion coefficient was obtained from the correlator of  vertex operators on the 2-dimensional worldsheet.
Indeed, the $1/S$ corrections of  \eqref{smallnuJ} are absent from (4.28) of \cite{Roiban:2010fe}.
The disagreement arises because in \cite{Roiban:2010fe} the approximate string solution was plugged in \eqref{3CPO} and this is not accurate enough to give the exact $\sigma$ integrals. Instead, we have kept the exact
solution and have expanded to get \eqref{smallnuJ} only after we have evaluated all integrals.

Another interesting limit is the

$\bullet$ Short string solution

In this case we set $\nu=0$ and $S/\sqrt{\lambda}<<1$. Then it easy to show \cite{Gubser:2002tv},\cite{Frolov:2002av} that
\begin{eqnarray}\label{smallS}
\frac{S}{\sqrt{\lambda}}=\frac{1}{2 w'^2},\,\frac{E}{\sqrt{\lambda}}=\frac{1}{w'} ,\,E^2=\sqrt{\lambda}\,2S,\,\,\,\,w'>>1.
\end{eqnarray}
Finally, for $w'>>1$ the fusion coefficient of \eqref{3CPO2} becomes
\begin{eqnarray}\label{smallSC}
C_{{\bar O}_{SJ}  O_{SJ} {\cal O}_I}=\frac{(l^2-1)\, \sqrt{l \lambda} \,\pi^{1/2}\,\Gamma(\frac{l}{2})}
{2^{l+3} \,N \,\Gamma(\frac{l+3}{2})} \sqrt{\frac{2S}{\sqrt{\lambda}}}=\frac{(l^2-1)\, \sqrt{l} \,\pi^{1/2}\,\Gamma(\frac{l}{2})}
{2^{l+3} \,N \,\Gamma(\frac{l+3}{2})} E.
\end{eqnarray}
Consequently, for the case of a short string the fusion coefficient scales as the square root of the spin
$\sqrt{2S\sqrt{\lambda}}=E$ and is proportional to the energy $E$ of the string.

\subsection{Evaluation of the correlator $<\bar{{\cal O}}_{SJJ}(w_3)\,{\cal O}_{SJJ}(w_1)\, {\cal O}_I(x)>$}

We now turn to the second correlation function \eqref{corr2} involving the massive string state with
three angular momenta $S,\,J_1,\,J_2=J_1$ and the BPS operator of \eqref{CPO} at strong coupling.
By plugging in \eqref{CPOstrong} the classical solution \eqref{PoincareJJ}
one gets
\begin{eqnarray}\label{3CPOJJ}
C_{{\bar O}_{SJJ}  O_{SJJ} {\cal O}_I}=\frac{(l+1) \sqrt{l \lambda}}{2^l \pi N}\int_{-\infty}^{\infty} d\tau \int_{0}^{\pi/2}d\sigma
\frac{e^{-l \nu \tau}\sin^l{n \sigma}}{\cosh^l{k \tau}\cosh^l{ \rho}}(\frac{\rho'^2}{\cosh^2{\rho}}
-k^2\tanh^2{k \tau}+\nu^2)\nonumber\\
\end{eqnarray}
when $l$ is even and zero when $l$ is odd.

In \eqref{3CPOJJ}
\begin{eqnarray}\label{Virasoronew1}
\rho'=\sqrt{k^2-\tilde{\nu}^2}\sqrt{\cosh^2{\rho}-w'^2\sinh^2{\rho}},\,\,w'^2=\frac{w^2-\tilde{\nu}^2}{k^2-\tilde{\nu}^2},\,
\tilde{\nu}^2=\nu^2+n^2
\end{eqnarray}
We see that the winding number $n$ enters \eqref{3CPOJJ} only through $\rho'$.
Using \eqref{Virasoronew1} one can rewrite \eqref{3CPOJJ} as
\begin{eqnarray}\label{3CPOJJ1}
C_{{\bar O}_{SJJ}  O_{SJJ} {\cal O}_I}=\frac{(l+1) \sqrt{l \lambda}}{2^l \pi N}\Big(\int_{-\infty}^{\infty} d\tau
\frac{e^{-l \nu \tau}}{\cosh^l{k \tau}}\int_{0}^{\rho_0}d\rho
\frac{\sqrt{k^2-\tilde{\nu}^2}\sqrt{\cosh^2{\rho}-w'^2\sinh^2{\rho}}\sin^l{n \sigma}}{\cosh^{l+2}{ \rho}}+\nonumber\\
\int_{-\infty}^{\infty} d\tau
\frac{e^{-l \nu \tau}(\nu^2-k^2\tanh^2{k\tau})}{\cosh^l{k \tau}}\int_{0}^{\rho_0}d\rho
\frac{\sin^l{n \sigma}}{\sqrt{k^2-\tilde{\nu}^2}\sqrt{\cosh^2{\rho}-w'^2\sinh^2{\rho}}\cosh^l{ \rho}}
\Big)\nonumber\\.
\end{eqnarray}
The $\sigma$ appearing in \eqref{3CPOJJ1} should be though as a function of $\rho$ which one gets when one integrates the first equation in \eqref{Virasoronew1}. This function is a hypergeometric one and we do not write it since it is not particularly illuminating.
Unfortunately this complicated dependence of $\sigma$ on $\rho$ make the $\sigma$ integrals of \eqref{3CPOJJ1} very difficult to evaluate. The $\tau$ integrals are the same as in the previous subsection.

However, one can estimate the behaviour of the fusion coefficient \eqref{3CPOJJ} in the large $S$ limit.
For this it is enough to notice that the denominators of the $\rho$ integrals in \eqref{3CPOJJ1} increase exponentially
with $\rho$. Taking into account that for large $S$, $\rho\approx \mu \sigma,\, \mu\approx \frac{1}{\pi}\log{S/\sqrt{\lambda}}>>1$ \cite{Frolov:2002av} we see that in the region where
the integrands are essentially non-zero one can approximate the sinus appearing in the numerators with
$\sin^l{n \sigma}\approx \Big(\frac{n \rho}{\mu}\Big)^l=\Big(\frac{n \pi \rho}{\log{S/\sqrt{\lambda}}}\Big)^l$, if $n=fixed$.
This approximation leads to the following behaviour of the fusion coefficient
\begin{eqnarray}\label{3CPOJJL}
C_{{\bar O}_{SJJ}  O_{SJJ} {\cal O}_I}\sim \frac{n^l \pi^l}{\log^l{\frac{S}{\sqrt{\lambda}}}}.
\end{eqnarray}
It is important to notice the very different scaling of the coefficient of \eqref{smallnuJ} and that of
\eqref{3CPOJJL}. In the first case (string solution with two angular momenta $S,\,J$) the coefficient
tends to a constant as $S\rightarrow \infty$ while in the second case (string solution with three angular momenta $S,\,J_1,\,J_2=J_1$) it scales as $1/\log^l{\frac{S}{\sqrt{\lambda}}}$.

\begin{figure}[!h]
  \centering
  \includegraphics[width=.5\textwidth]{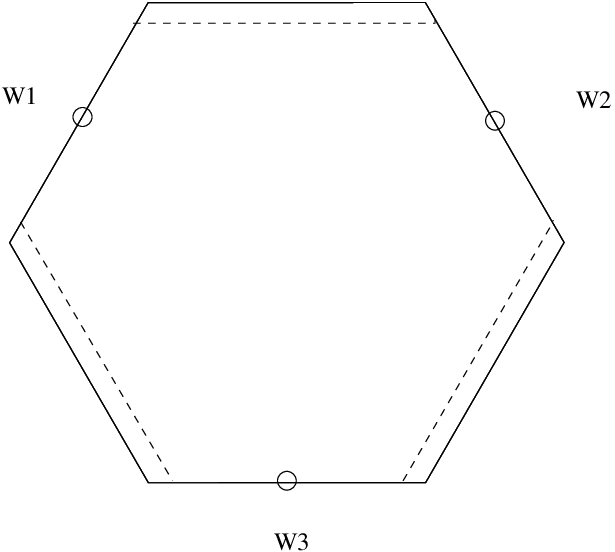}
  \caption{Projection on the $y^1-y^2$ plane of the 6-gluon scattering amplitude solution.
  $X_i,\,i=1,...,6$ are the positions of the cusps.
  The dashed lines denote
  the tips of the strings which almost touch the boundary when the spins $S_i \rightarrow \infty,\, i=1,2,3$.
  The string which lies on the boundary from $x_1$ to $x_2$  splits into two strings
  which end on the boundary again extending from $x_3$ to $x_4$ and $x_5$ to $x_6$, respectively.
  The three twist 2 operators
  will be situated are at $w_1$, $w_2$ and $w_3$ which are the centers of mass of the three strings when these are on the boundary. }
  \label{fig:Figure2}
\end{figure}
We would like to close this Section by commenting on the possibility of relating
the 3-point function coefficient of three twist 2 operators with very large spin $S/\sqrt{\lambda}>>1$, at strong coupling,
to the action of the  6-gluon scattering amplitude of Figure 2.
In Section 2.1, we have commented on the similarity between the 4-gluon scattering amplitude and
a certain analytic continuation
of the GKP solution.
This similarity is valid in the infinite spin limit \footnote{For another study of the large spin limit of operators in the $SL(2)$ sector see \cite{me-Savv}.} where the string solution describing
the tunneling of a string from boundary to boundary does become the solution
describing the 4-gluon scattering amplitude.
In fact, one can get the 2-point correlators of Section 2, for the cases of two large spin S operators by subtracting
the action of the two small strips of Figure 1b from the action of the 4-gluon scattering amplitude of Figure 1.

Thus, it may be possible that one can use the known action for the 6-gluon scattering amplitude to
extract information about the 3-point correlator of three large spin twist 2 operators.
Again in the limit where the spins $S_i\rightarrow \infty,\, i=1,2,3$ the tips of each string touch the boundary and
a light-like Wilson loop with 6 edges emerges.
One can view the minimal surface of Figure 2 as the solution for a folded string,  sitting on the boundary
from almost $x_1$ to almost $x_2$, that propagates in the bulk and splits
into two strings that end on the boundary again, from  $x_3$ to $x_4$
and $x_5$ to $x_6$, respectively. One would have to
subtract the area of the three strips of Figure 2 taking this way into account
the fact that the corresponding operators have finite spins $S_i,\, i=1,2,3$.
The cross ratios of the scattering amplitude should be, somehow, related to
dot products of  the three light-like vectors defining the three directions along which we take the covariant derivatives in the three operators.
It would be nice to see if this possibility can be realised or not.
Finally, one could  hope that the correlators of light-like Wilson lines in $N=4$ SYM
chopped at the end as in Figure 1b and Figure 2 would be related to
correlators of large spin twist 2 operators even at weak coupling, as it happens with the scattering amplitudes
\cite{Drummond:2007aua,Brandhuber:2007yx}. It would also be interesting to analyse the behaviour of the 3-point functions
in the limit where the corresponding Wilson loop develops a self-crossing \cite{Georgiou:2009mp}.

\appendix

\section{Appendix}

In this appendix we give the expansions of the hypergeometric functions needed to derive equations \eqref{smallnuJ} and \eqref{smallSC}.
The integrals of Section 3 have the following expansions around $w'=1$
\begin{eqnarray}\label{it1approx}
I_{\tau 1} = \frac{\sqrt{\pi} \Gamma(\frac{l}{2})}{k \Gamma(\frac{1 + l}{2})}+O(\nu),
\end{eqnarray}
\begin{eqnarray}\label{it2approx}
I_{\tau 2} = \frac{\sqrt{\pi} \Gamma(\frac{l}{2}+1)}{k \Gamma(\frac{3 + l}{2})}+O(\nu),
\end{eqnarray}
\begin{eqnarray}\label{is1approx}
I_{\sigma 1} = \frac{k l \sqrt{\pi} \Gamma(\frac{l}{2})}{4 \Gamma(\frac{3 + l}{2})} - \frac{
  k \sqrt{\pi}  \Gamma(\frac{l}{2})}{4 \Gamma(\frac{3 + l}{2})}(w'-1)+O((w'-1)^{3/2})
\end{eqnarray}
\begin{eqnarray}\label{is2approx}
I_{\sigma 2} = 2^{l - 2} \frac{\Gamma^2(\frac{l}{2})}{k \Gamma(l)} - 2^{l - 2} \frac{\Gamma^2(\frac{l}{2})}{2 k (1 - \frac{l}{2}) \Gamma(l)}(w' - 1)-
\frac{2^{\frac{3 l}{2}-1}\Gamma^2(\frac{l}{2})\Gamma(1-\frac{l}{2})\Gamma(\frac{1+l}{2})}{l k \Gamma(\frac{1-l}{2})\Gamma(l)\Gamma(\frac{l}{2})}
(w'-1)^{\frac{l}{2}}\nonumber\\,
\end{eqnarray}
where the last term in \eqref{is2approx} is not needed for $l>2$.
For the case of short string, $w'>>1$, the relevant expansions are
\begin{eqnarray}\label{is1approxsmallS}
I_{\sigma 1} = \frac{k \pi}{4 w'}+O(1/w'^2)
\end{eqnarray}
and
\begin{eqnarray}\label{is1approxsmallS}
I_{\sigma 2} =\frac{ \pi}{2k w'}+O(1/w'^2).
\end{eqnarray}

\vspace{2cm}

\noindent {\large {\bf Acknowledgments}}

\vspace{3mm}

\noindent
We wish to thank Dimitrios Giataganas, Valeria Gili, Rodolfo Russo, George Savvidy
and Dimitrios Zoakos for useful discussions and comments.

\end{document}